\documentclass[conference]{IEEEtran}
\IEEEoverridecommandlockouts
\usepackage{cite}
\usepackage{amsmath,amssymb,amsfonts}
\usepackage{amsthm}
\theoremstyle{definition}

\usepackage{algorithmic}
\usepackage{graphicx}
\usepackage{textcomp}
\usepackage{xcolor}
\usepackage{tcolorbox}
\usepackage{tabularx}
\usepackage{booktabs}
\usepackage{float}
\usepackage{listings}
\usepackage{xcolor}
\usepackage{enumitem}
\usepackage{tikz}
\usepackage{placeins}
\usepackage{stfloats}
\usepackage[normalem]{ulem}
\usepackage[hidelinks]{hyperref}

\def\BibTeX{{\rm B\kern-.05em{\sc i\kern-.025em b}\kern-.08em
    T\kern-.1667em\lower.7ex\hbox{E}\kern-.125emX}}
\begin{document}

\title{{ProvICS: A Provenance-based Intrusion Detection for Industrial Control Systems}\\
\thanks{This work is supported by the National Science Foundation, Award \# 2239609.}
}

\author{\IEEEauthorblockN{Md Neyamul Islam Shibbir}
\IEEEauthorblockA{\textit{Department of Computer Science} \\
\textit{The University of Texas at El Paso}\\
mshibbir@miners.utep.edu}
\and
\IEEEauthorblockN{Deepak K Tosh}
\IEEEauthorblockA{\textit{Department of Computer Science} \\
\textit{The University of Texas at El Paso}\\
dktosh@utep.edu}

}

\maketitle

\begin{abstract}
The convergence of Information Technology and Operational Technology has exposed Industrial Control Systems (ICS) to multi-stage cyberattacks that traverse software, network, and physical process layers simultaneously. Although Provenance-based Intrusion Detection Systems (PIDS) are effective in Information Technology (IT) environments, their applicability to Industrial Cyber-Physical Systems (CPS) remains largely unexplored because of the absence of datasets that jointly capture host-level causal behavior, industrial network semantics, and physical process state. To address this gap, we design an open-source, Hardware-in-the-Loop (HIL) CPS testbed that replicates an industrial chemical reactor control architecture across the Purdue model layers. Using this testbed, we propose ProvICS, a multimodal provenance dataset purpose-built for CPS intrusion detection, which synchronously captures four streams: whole-system provenance graphs from the supervisory host and the resource-constrained PLC, decoded Modbus deep-packet inspection records, and physical process telemetry. The collection comprises a 48-hour benign phase and a 22-hour attack phase across four campaigns covering 20 ICS ATT\&CK techniques over 32 attack events, ranging from reconnaissance to physical process manipulation. Comparative analysis shows that ProvICS is among the few existing ICS/CPS benchmarks with multi-host kernel-level provenance, real PLC hardware-in-the-loop execution, decoded Modbus traffic, physical process-state measurements, and auxiliary raw PCAP traces in a time-synchronized collection. Baseline detection further confirms that cross-modal fusion can detect all 32 labeled attack events (F1 = 0.913, false-positive rate (FPR) = 1.40\%), demonstrating the dataset's ability to expose complementary attack signals across modalities and addressing a gap not covered by prior benchmarks.
\end{abstract}

\begin{IEEEkeywords}
Operational Technology, Provenance-based Intrusion Detection, Multimodal Dataset, Hardware-in-the-Loop Testbed, ICS Security
\end{IEEEkeywords}

\section{Introduction}
Critical infrastructures including power grids, industrial manufacturing, water systems, healthcare, and transportation networks have increasingly adopted computational and digital technologies. While these advancements have substantially improved operational efficiency, they simultaneously expand the attack surface of systems whose compromise can carry catastrophic consequences. A successful cyberattack against such infrastructure can trigger significant physical incidents, threaten national security, endanger public safety, and cause severe disruptions to essential services~\cite{roumani2025examining}. Modern Industrial Control System (ICS) attacks span host, network, and physical layers, making them impossible to reconstruct from a single viewpoint. Data provenance solves this by tracking kernel-level causal relationships, linking seemingly benign events across the entire control stack to reveal full, cross-layer attack paths that isolated monitoring misses~\cite{li2023nodlink}. Intrusion Detection Systems (IDSs) have been extensively studied in IT environments, yet they exhibit fundamental limitations against modern, multi-stage threats~\cite{li2023nodlink}. These limitations are further compounded in OT contexts, where conventional IDS solutions are designed predominantly for IT-layer monitoring which can lack the visibility necessary to account for process-based behaviors and physical state dynamics central to Industrial Control System (ICS) security~\cite{anthi2021three, wani2025kids}.

A fundamental barrier to IDS research in the OT domain is the lack of open-source resources. Industrial deployments are prohibitively expensive, and most factory components are proprietary systems governed by vendors that rely on ``security through obscurity'' \cite{formby2018lowering} \cite{silverman2020study}. Vendors rarely disclose architectural details, communication protocols, or internal configurations, a posture that, despite its prevalence, is no longer considered a viable defense strategy~\cite{stouffer2011guide}. This inaccessibility severely constrains reproducible security research on live ICS infrastructure. 
Furthermore, effective IDS development requires telemetry that 
captures every observable dimension of the CPS environment, including 
host-level system provenance across all nodes, industrial network traffic, and 
physical process states. Partial observability risks missed attack signals and 
an incomplete ground truth for evaluation~\cite{shin2020hai}.

In computer security, data provenance graphs have emerged as a powerful 
paradigm for intrusion detection and forensic 
investigation~\cite{jiang2025orthrus}. These graphs capture causal 
relationships among system entities such as processes, files, and network connections by 
using information derived from audit logs. By linking events and their 
dependencies, data provenance enables full reconstruction of an attack 
sequence, revealing insights that isolated log analysis or network flow 
inspection alone cannot provide~\cite{bilot2025sometimes,wang2024incorporating}. 
While Provenance-based Intrusion Detection Systems (PIDS) have demonstrated 
strong efficacy in traditional IT environments, their applicability to OT/ICS 
settings where attacks manifest concurrently across software, industrial 
communication protocols, and physical processes remains largely unexplored.

These gaps motivate the following research questions:
\begin{itemize}
  \item Given the resource-constrained nature of CPS, how can we collect system telemetry, and network packets for enabling real-time provenance-based intrusion detection?
  \item How can provenance-based multimodal data serve as an effective foundation for intrusion detection in OT environments?
\end{itemize}

To address these questions, this paper makes the following contributions:

\begin{itemize}
  \item We develop an open-source and lightweight CPS testbed that architecturally replicates the behavioral characteristics of a standard industrial control systems environment.
  \item We outline a systematic, multimodal data collection process that 
  captures host-level system provenance, industrial network traffic, and 
  physical process telemetry from ICS environments.
  \item To the best of our knowledge, this work is among the first to provide an 
  open-source multimodal provenance dataset designed specifically for CPS 
  intrusion detection, which is publicly available on Hugging 
  Face.\footnote{\textcolor{blue}{\url{https://huggingface.co/datasets/trucyberlab/multimodal-ICS-provenance}}}
\end{itemize}

\section{Related Work}
As previously mentioned, there is a significant lack of research on provenance-based intrusion detection within Cyber-Physical Systems (CPS) environments. Consequently, there is a significant shortage of provenance-based intrusion detection datasets available for study. Ghiasvand et al. \cite{ghiasvand2024cicapt} proposed the CICAPT-IIoT dataset, a provenance-based Advanced Persistent Threat (APT) dataset specifically designed for Industrial IoT (IIoT) settings. However, they only collected system provenance data from a single node despite the testbed containing multiple virtual machines, Raspberry Pis, and physical sensors. This restricts the dataset's applicability for analyzing distributed, network-wide APT activities, particularly lateral movement across diverse hosts within a CPS architecture. Furthermore, a critical examination of their published dataset reveals a significant gap in data-flow coverage. The provenance graphs record process activity and file/socket events such as opening files or creating connections, but they largely miss actual data-transfer operations like read, write, sendto, and recvfrom. Because of this, the graphs show who connected to what, but not how data actually moved. This makes it harder to detect behavior such as real data transfer, lateral movement, or data exfiltration accurately.
To our knowledge, no other existing CPS intrusion detection dataset includes system provenance data. While several existing datasets capture network traffic and physical state information, they frequently lack the depth necessary to facilitate the detection of Advanced Persistent Threats (APTs). For example, Mathur et al. \cite{mathur2016swat} proposed the SWaT dataset, which represents a scaled-down, high-fidelity replica of a modern six-stage water treatment facility. The data collected include network traffic (PCAPs) and values from 51 sensors and actuators. Another testbed \cite{morris2011control} combined model control systems from multiple critical infrastructure industries, such as power and water, to provide a realistic environment for security research and training. However, while it effectively demonstrates common industrial protocols and vulnerabilities, it primarily focuses on network-level data and lacks the internal system provenance needed for detecting advanced, multi-stage APT threats. Another dataset named ICS-Flow \cite{dehlaghi2023anomaly} provides integrated network traffic and process state logs from simulated industrial components to support both supervised and unsupervised machine learning. While it covers common network attacks, it lacks the deep system provenance data required to track the internal logic of APTs. In our work, we propose a complete dataset that collects system information from all the components of the environment; in addition, it includes network information and the physical state of the system at any given time.

\section{Provenance-aware OT architecture}

In this work, we present an architecture that has the capability to capture provenance information from heterogeneous ICS components by instantiating the sensing-actuation feedback loop across the physical, control, and supervisory layers. This architecture includes a dedicated PLC that sits at Purdue Level 1 and executes a deterministic scan-cycle loop that repeatedly reads sensor feedback values from a physical plant (Purdue Level 0), which is a digital twin~\cite{fuller2020digital} of a continuous stirred-tank chemical reactor, runs control logic to compute actuator setpoints, and writes those outputs back to the plant, thereby closing the sensing and actuation feedback loop that maintains the chemical reactor at its target operating state. A SCADA HMI (Purdue Level 2) serves as the supervisory boundary between the human operator and the automated control loop by continuously polling all mapped process variables from the PLC, rendering the live plant state as operator-interpretable visualizations, and translating operator decisions into setpoint write commands that propagate downward through the architecture to effect physical change in the plant. Furthermore, a Historian (Purdue Level 3) functions as the passive, read-only time-series archive of the OT architecture, continuously storing all process-variable tags polled by the SCADA layer and compressing and persisting each timestamped observation into a long-term site-wide database without issuing any write commands to the control plane. 

\section{Modeling Multimodal Data Relationship}
We formalize ProvICS as a multimodal observation space $\mathcal{D} = \{M_1, M_2, M_3, M_4\}$, where each modality $M_i$ captures a complementary projection of system behavior. Modalities are distinguished by their generating process and representational structure rather than by sensory format.

\begin{figure}[t]
\centering
\includegraphics[width=0.9\columnwidth]{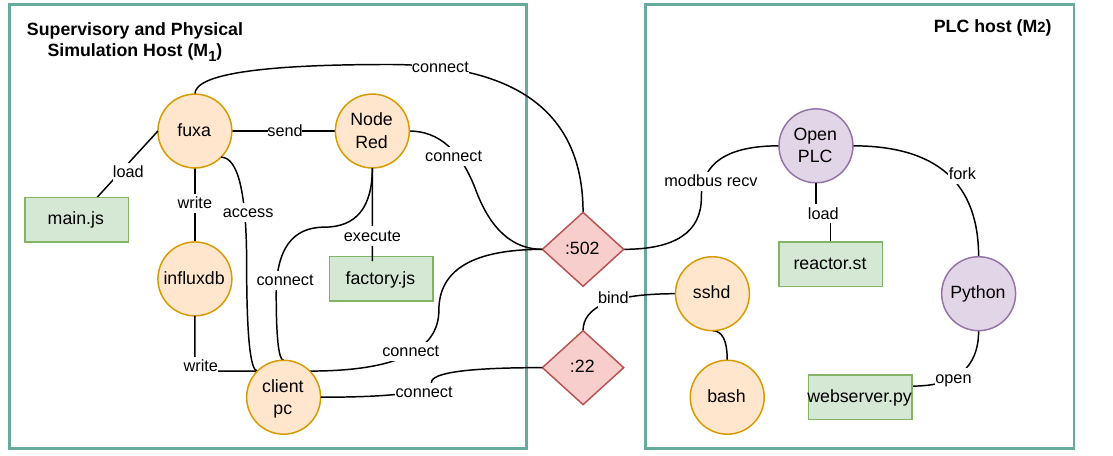}
\caption{Provenance Graph Representation}
\label{fig:provenance_workflow}
\vspace{-1.8em}
\end{figure}

\textbf{Modality $M_1$: Host Provenance Graph.} A directed acyclic graph $G_h = (V_h, E_h)$ capturing kernel-level causal dependencies on the supervisory host, where $V_h = V_p \cup V_f \cup V_n$ represents process($V_p$), file($V_f$), and network socket($V_n$) entities, and each edge $e \in E_h$ is annotated with a syscall operation and timestamp: $e = (v_i, v_j, op, t)$.

\textbf{Modality $M_2$: PLC Provenance Graph.} A graph $G_{plc} = (V_{plc}, E_{plc})$ analogous to host graph $G_h$ but captured on the PLC host, augmented with scan-cycle instrumentation edges that expose internal control logic decisions, bridging the otherwise opaque boundary between network inputs and physical outputs. We can see provenance graph samples for modalities $M_1$ and $M_2$ in Figure~\ref{fig:provenance_workflow}.

\textbf{Modality $M_3$: Protocol Semantic Capture.} A sequence of application-layer records $S = \{s_1, s_2, \ldots, s_n\}$, where each $s_i = (t_i, fc_i, addr_i, val_i)$ encodes the timestamp, function code, register address, and payload of an ICS protocol transaction (e.g., Modbus/TCP).

\textbf{Modality $M_4$: Physical Process State.} A multivariate time-series $\mathbf{X}(t) = [x_1(t), x_2(t), \ldots, x_k(t)]^T$ of $k$ process variables sampled at frequency $f_s$, representing the plant's dynamic response to both legitimate control actions and adversarial manipulations.

From here, researchers have a variety of options for using this multimodal data, including cross-modality correlation, score-level fusion, and causal reconstruction. Synchronized timestamps and socket identities enable events from provenance, protocol, and physical-process streams to be fused into unified causal chains, while modality-specific anomaly scores can be combined to improve detection coverage across heterogeneous attack phases.

\section{Data Collection Infrastructure}

\subsection{Testbed Configuration}

The proposed testbed comprises two physical computation nodes. There is a Raspberry Pi 4 Model B~\cite{raspberrypi_website} (1\,GB 
RAM) running Debian 11 (64-bit), hosting the OpenPLC Runtime v3 \cite{alves2018openplc} as the PLC. The second is an x86-64 workstation (16 GB RAM) running Ubuntu 22.04 LTS~\cite{ubuntu}, hosting the SCADA stack, the physical plant simulator, 
and the data collection infrastructure. 

\subsubsection{Network Emulation Environment (CORE Emulator)}
\begin{figure}[t]
\centering
\includegraphics[width=0.7\columnwidth]{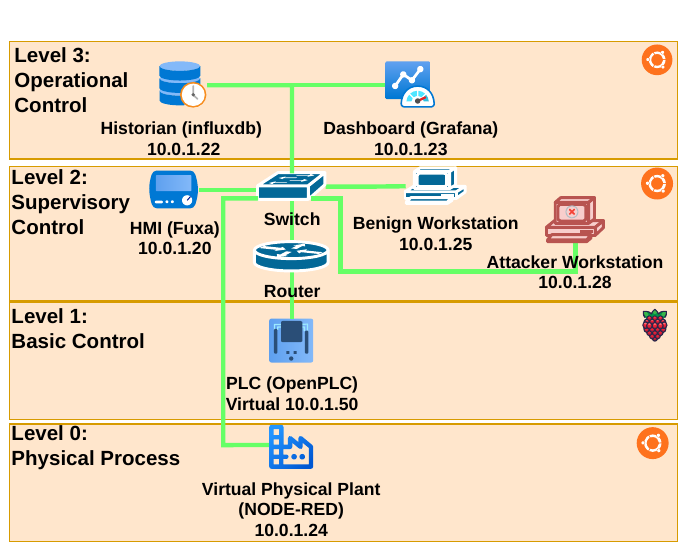}
\caption{Our proposed CPS testbed with the Purdue reference model}
\label{fig:architecture_network}
\vspace{-2.4em}
\end{figure}

The CORE network emulator~\cite{ahrenholz2008core} 
provides an isolated virtual network with dedicated per-node IP 
addressing with \texttt{10.0.1.0/24} subnet of the testbed. All virtual nodes interconnect through a central CORE 
router node. Connectivity between the CORE virtual network and the 
physical Raspberry Pi is established via a \texttt{veth-bridge} 
interface, which bridges the emulated network to the physical 
host network adapter. Modbus/TCP traffic destined for the 
virtual controller address \texttt{10.0.1.50:502} is 
transparently forwarded to the physical OpenPLC instance at 
\texttt{Physical\_IP:502} via DNAT rule. Table~\ref{tab:network} summarizes the network 
configuration. In addition, Figure~\ref{fig:architecture_network} shows the network topology of the environment.

\begin{table}[h]
\centering
\caption{Testbed Network Configuration}
\label{tab:network}
\footnotesize
\setlength{\tabcolsep}{3pt}
\begin{tabularx}{\columnwidth}{>{\raggedright\arraybackslash}p{0.26\columnwidth} >{\raggedright\arraybackslash}p{0.14\columnwidth} >{\raggedright\arraybackslash}p{0.22\columnwidth} >{\raggedright\arraybackslash}p{0.18\columnwidth} >{\raggedright\arraybackslash}X}
\hline
\textbf{Node} & \textbf{UID} & \textbf{IP Address} & \textbf{Role} & \textbf{Key Ports} \\
\hline
Raspberry Pi 4 (OpenPLC) & 0 & 10.0.1.50 & PLC & 502,20,  8080\\
Node-RED & 1101 & 10.0.1.24 & Plant Simulator & 1880 \\
FUXA & 1100 & 10.0.1.20 & HMI & 1881 \\
InfluxDB & 1102 & 10.0.1.22 & Historian & 8086 \\
Grafana & 1103 & 10.0.1.23 & Visualization & 3000 \\
Kali Linux & 1105 & 10.0.1.28 & Attack Node & -- \\
Ubuntu 22.04 & 1104 & 10.0.1.25 & Benign Node & --\\
\end{tabularx}
\vspace{-1.5em}
\end{table}

\subsubsection{Programmable Logic Controller (PLC)}

The OpenPLC Runtime v3~\cite{alves2018openplc} executes an IEC 61131-3 Structured Text (\texttt{.st}) control program that implements the chemical reactor's process logic. It is deployed on the Raspberry Pi (\texttt{10.0.1.50}); this embedded device serves as the physical Hardware-in-the-Loop (HIL) Programmable Logic Controller (PLC) within our testbed.

\subsubsection{Physical Plant Simulator}
The simulated physical plant is modelled after a continuous stirred-tank reactor (CSTR) with gas–liquid separation, inspired by the Tennessee Eastman (TE) challenge process~\cite{downs1993plant}, a widely adopted benchmark in process control and fault-detection research~\cite{ekisa2022vicsort}. The plant is implemented in \textbf{Node-RED}~\cite{ferencz2019using} as a software-based digital twin, hosted at \texttt{10.0.1.24:1880}, where it executes the process dynamics in simulation and exchanges sensor readings and actuator commands with the PLC exclusively via the Modbus/TCP protocol.

\subsubsection{Supervisory SCADA Stack}

The supervisory stack comprises two Dockerized services 
managed within the CORE emulation environment:

\textbf{1. HMI (FUXA~\cite{fuxa_documentation_2026})} (\texttt{10.0.1.20}, port 1881) FUXA performs cyclic register polling via FC3 at one-second intervals 
    across all mapped registers and issues FC16 commands 
    for operator setpoint changes.

\textbf{2. Historian (InfluxDB~\cite{influxdb})} (\texttt{10.0.1.22}, port 8086) Time-series historian. FUXA writes all polled process 
    variables at one-second resolution, producing a continuous 
    multivariate telemetry record across all the variables summarized in Table \ref{tab:data_dictionary}

\subsection{Collection Attributes for ProvICS}
\textbf{Provenance Data Collection ($M_1$):}
A daemon service (Auditd~\cite{audituserspace}) is used on the supervisory host to capture the whole-system provenance, which is inspired by the SPADE framework~\cite{gehani2012spade}.

\textbf{PLC Provenance Data Collection ($M_2$):} On the Raspberry Pi, we configured targeted \texttt{auditd} rules on the controller to record PLC-specific system calls. These audit logs are subsequently parsed and translated into a provenance graph.

\textbf{Protocol Semantic Capture ($M_3$):} At the analytical level, a 
concurrent \texttt{tshark} dissection stream performs real-time 
deep-packet inspection of Modbus/TCP traffic, extracting 
structured protocol fields such as function code, register address, 
word count, and payload exported into machine-readable JSONL records which provides the ground-truth evidence of \emph{what} was written, \emph{where}, and \emph{when} records that can be directly correlated with provenance graph edges to reconstruct the causal chain from attacker action to physical consequence.

\textbf{Physical Process States Data Collection ($M_4$):} Physical process state data is collected through FUXA's built-in historian integration, which polls all mapped Modbus holding registers from OpenPLC at one-second intervals via FC3 Read Holding Registers requests and writes the resulting values directly to InfluxDB.

Data collection is organized into two runs: a \textit{benign run} capturing 48 hours of normal operational behavior with no adversarial activity, and an \textit{attack run} executing four adversarial campaigns (C1--C4) comprising 32 attack events across 27 labeled phases over 22 hours, with all collection services active throughout. All components shared a common UTC time reference via NTP.

\begin{table*}[!t]
\centering
\caption{Dataset Properties: Variables and Attributes Across All Modalities}
\label{tab:data_dictionary}
\scriptsize
\setlength{\tabcolsep}{3.2pt}
\renewcommand{\arraystretch}{0.85}
\begin{tabularx}{\textwidth}{@{}>{\raggedright\arraybackslash}p{2.7cm}>{\raggedright\arraybackslash}p{6.2cm}X@{}}
\toprule
\textbf{Modality} & \textbf{Variables} & \textbf{Description} \\
\midrule
\multicolumn{3}{@{}l@{}}{\textit{Physical Process State ($M_4$)}} \\
\cmidrule(l){1-3}
Operator Setpoints &
\texttt{flow\_set}, \texttt{a\_setpoint}, \texttt{pressure\_sp}, \texttt{level\_sp}, \texttt{override\_sp} &
HMI-issued reference values governing the reactor operating point. \\
PLC Setpoints &
\texttt{f1\_valve\_sp}, \texttt{f2\_valve\_sp}, \texttt{purge\_valve\_sp}, \texttt{Product Valve SP} &
Valve positions computed by the PLC control logic from operator setpoints. \\
Process Variables &
\texttt{Pressure}, \texttt{LevelPV}, \texttt{F1 Flow}, \texttt{f2\_flow}, \texttt{purge\_flow}, \texttt{ProductFlowPV}, \texttt{A in purge PV}, \texttt{b\_in\_purge}, \texttt{c\_in\_purge}, \texttt{Product} &
Sensor telemetry from the continuous reactor, including pressure, level, flow, and composition measurements. \\
Actuator Feedback &
\texttt{F1 Valve Position Feedback}, \texttt{F2 valve Postion Feedback}, \texttt{Purge Valve Feedback} &
Physical valve position feedback confirming actuator state. \\
\addlinespace[0.2em]
\multicolumn{3}{@{}l@{}}{\textit{Protocol Semantic Network ($M_3$)}} \\
\cmidrule(l){1-3}
Flow Identity &
\texttt{ip\_src}, \texttt{ip\_dst}, \texttt{tcp\_srcport}, \texttt{tcp\_dstport} &
IP/port tuples defining IT--OT communication channels. \\
Modbus Application &
\texttt{modbus\_func\_code}, \texttt{modbus\_reference\_num}, \texttt{modbus\_data} &
Function code, register address, and raw payload of each Modbus transaction. \\
Timing &
\texttt{frame\_time\_epoch} &
Packet timestamp for inter-arrival analysis and polling-frequency baselines. \\
\addlinespace[0.2em]
\multicolumn{3}{@{}l@{}}{\textit{Host Provenance ($M_1$)}} \\
\cmidrule(l){1-3}
Process Vertices &
\texttt{type}, \texttt{name}, \texttt{exe}, \texttt{command line}, \texttt{cwd}, \texttt{pid}, \texttt{ppid}, \texttt{tgid}, \texttt{uid}, \texttt{euid}, \texttt{gid}, \texttt{egid}, \texttt{seen time}, \texttt{start time}, \texttt{source} &
Execution context of supervisory services such as Node-RED, FUXA HMI, InfluxDB, and Grafana. \\
Artifact Vertices &
\texttt{subtype}\,\textsuperscript{a}, \texttt{path}, \texttt{permissions}, \texttt{version}, \texttt{epoch}, \texttt{fd}, \texttt{read fd}, \texttt{write fd} &
File, pipe, and device descriptors accessed by host processes. \\
Causal Edges &
\texttt{type}, \texttt{operation}\,\textsuperscript{b}, \texttt{time}, \texttt{event id}, \texttt{pid}, \texttt{size}, \texttt{flags}, \texttt{mode}, \texttt{source} &
Syscall-level causal links with payload size and permission flags. \\
\addlinespace[0.2em]
\multicolumn{3}{@{}l@{}}{\textit{PLC Provenance ($M_2$)}} \\
\cmidrule(l){1-3}
Process Vertices &
\texttt{type}, \texttt{name}, \texttt{exe}, \texttt{command line}, \texttt{pid}, \texttt{ppid}, \texttt{tgid}, \texttt{uid}, \texttt{euid}, \texttt{gid}, \texttt{egid}, \texttt{machine}, \texttt{note}, \texttt{seen time}, \texttt{start time}, \texttt{source} &
Execution context of the PLC runtime and its supporting services on the edge device. \\
Artifact Vertices &
\texttt{subtype}\,\textsuperscript{c}, \texttt{path}, \texttt{permissions}, \texttt{version}, \texttt{epoch}, \texttt{remote address}, \texttt{remote port}, \texttt{fd} &
Files, network sockets, and descriptors. \\
Causal Edges &
\texttt{type}, \texttt{operation}\,\textsuperscript{d}, \texttt{time}, \texttt{event id}, \texttt{size}, \texttt{flags}, \texttt{source} &
Syscall-level links; unlike the host provenance, these edges omit \texttt{mode} and \texttt{pid}. \\
\bottomrule
\end{tabularx}

\vspace{0.35em}
\parbox{\textwidth}{\footnotesize
\textsuperscript{a} Host subtypes: \texttt{file}, \texttt{directory}, \texttt{character device}, \texttt{eventfd}, \texttt{unnamed pipe}, \texttt{unknown}.\\
\textsuperscript{b} Host operations: \texttt{read}, \texttt{write}, \texttt{open}, \texttt{clone}, \texttt{fork}, \texttt{execve}, \texttt{exit}, \texttt{load}, \texttt{create}, \texttt{chmod}, \texttt{setuid}, \texttt{setgid}, \texttt{update}.\\
\textsuperscript{c} RPi subtypes: \texttt{file}, \texttt{directory}, \texttt{network socket}, \texttt{unknown}.\\
\textsuperscript{d} RPi operations: \texttt{read}, \texttt{write}, \texttt{open}, \texttt{connect}, \texttt{accept}, \texttt{bind}, \texttt{send}, \texttt{recv}, \texttt{fork}, \texttt{execve}, \texttt{exit}, \texttt{load}.}
\vspace{-1.5em}
\end{table*}

\section{Adversary Simulation}
\subsection{Threat Model}

We consider an adversary whose objective is to disrupt or destroy the physical process governed by the ICS, consistent with NIST SP 800-82~\cite{stouffer2023guide} and MITRE ATT\&CK for ICS~\cite{alexander2020mitre}. The adversary possesses network-level access, knowledge of standard industrial protocols, and multi-stage campaign capabilities spanning reconnaissance, lateral movement, and physical impact, consistent with documented threat actors such as those behind Stuxnet~\cite{farwell2011stuxnet} and CRASHOVERRIDE~\cite{slowik2018anatomy}.

The attack surface spans three layers. At the \textit{network layer}, the adversary may scan, enumerate protocol register spaces, intercept traffic, and flood control channels. At the \textit{host layer}, the adversary may exploit remote services to gain execution, escalate privileges, establish persistence, and move laterally. At the \textit{physical layer}, the adversary may inject false register values, manipulate setpoints, upload modified control logic, or spoof sensor feedback to induce unsafe states while evading process-level alarms.

\subsection{Attack Campaigns}
To generate labelled attack data, four campaigns were executed against the live CPS testbed from a dedicated Kali Linux~\cite{kali_docs} node, targeting the OpenPLC controller, Node-RED physical simulator, FUXA HMI, and InfluxDB historian. Each campaign ran as an autonomous Python script producing a ground-truth CSV with UTC-timestamped phase boundaries, ICS ATT\&CK mappings, and action descriptions. The campaigns span contrasting detection profiles, from aggressive, network-noisy intrusions to persistent threats generating minimal attacker-originated traffic.

\textbf{Campaign 1 (C1): Smash and Grab.} A noise-indifferent adversary performs high-rate scanning, Modbus enumeration, setpoint manipulation, unauthorised actuation, sensor spoofing, and multi-layer denial-of-service flooding before restoring all values. Serves as the high-visibility reference case.

\textbf{Campaign 2 (C2): Low and Slow APT.} A stealth-oriented adversary uses passive observation and low-rate enumeration mimicking legitimate polling, then gradually drifts process parameters, establishes an adversary-in-the-middle position, and tampers with historian records.

\textbf{Campaign 3 (C3): Targeted Sabotage.} An insider adversary replaces the PLC Structured Text program with a malicious version that removes safety protections and destabilises control loops, while spoofing sensor feedback. Following impact, the attacker restores the original ST program and wipes historian records covering the anomaly window to eliminate forensic evidence. Decisive actions manifest exclusively as host-side program events rather than anomalous network traffic, making this campaign particularly relevant for provenance-based detection.

\textbf{Campaign 4 (C4): Full Spectrum Persistent Threat.} The most comprehensive scenario, combining persistence implantation, lateral movement, traffic interception, process manipulation, and historian tampering. After silent withdrawal, implanted PLC logic continues driving unsafe conditions without attacker-originated traffic, a phase uniquely valuable for evaluating detectors reliant on physical state or provenance traces rather than packet-level activity.

\begin{table}[htbp]
\centering
\caption{ICS ATT\&CK~\cite{alexander2020mitre} Technique Coverage Across Campaigns}
\label{tab:attack_coverage}
\scriptsize
\setlength{\tabcolsep}{5pt}
\renewcommand{\arraystretch}{1}
\begin{tabular}{llcccc}
\toprule
\textbf{ID} & \textbf{Technique Name} & \textbf{C1} & \textbf{C2} & \textbf{C3} & \textbf{C4} \\
\midrule
T0842 & Network Sniffing &  & \checkmark &  &  \\
T0846 & Remote System Discovery & \checkmark &  &  & \checkmark \\
T0861 & Point \& Tag Identification & \checkmark & \checkmark &  &  \\
T0859 & Valid Accounts & \checkmark & \checkmark & \checkmark & \checkmark \\
T0836 & Modify Parameter & \checkmark & \checkmark &  &  \\
T0855 & Unauthorized Command Message & \checkmark &  &  &  \\
T0856 & Spoof Reporting Message & \checkmark &  & \checkmark & \checkmark \\
T0814 & Denial of Service & \checkmark &  &  &  \\
T0807 & Command-Line Interface &  & \checkmark & \checkmark &  \\
T0801 & Monitor Process State &  & \checkmark & \checkmark &  \\
T0830 & Adversary-in-the-Middle &  & \checkmark &  & \checkmark \\
T0893 & Data from Local System &  & \checkmark &  & \checkmark \\
T0832 & Manipulation of View &  & \checkmark &  & \checkmark \\
T0809 & Data Destruction &  & \checkmark &  & \checkmark \\
T0886 & Remote Services &  &  & \checkmark & \checkmark \\
T0845 & Program Upload &  &  & \checkmark &  \\
T0889 & Modify Program &  &  & \checkmark & \checkmark \\
T0843 & Program Download &  &  & \checkmark &  \\
T0872 & Indicator Removal on Host &  &  & \checkmark & \checkmark \\
T0831 & Manipulation of Control &  &  &  & \checkmark \\
\bottomrule
\end{tabular}
\vspace{-3.0em}
\end{table}
Table \ref{tab:attack_coverage} summarises the ICS ATT\&CK technique coverage across campaigns. It comprises four adversarial campaigns with 32 attack events covering
20 unique ICS ATT\&CK techniques across 37 labeled technique-campaign pairs which
exceeds total number of events (32) because a single event may exercise multiple techniques
simultaneously. Taken together, the four campaigns exercise all four data collection modalities and include at least one attack stage that would be weakly observable or completely missed without each of them. This makes ProvICS suitable for evaluating provenance-based intrusion detection while also supporting broader multi-modal CPS security analysis.

\section{Dataset Assessment}

The ProvICS dataset spans four modalities, whose variables and attributes are 
fully enumerated in Table~\ref{tab:data_dictionary}. The 
physical process state ($M_4$) modality records 7 operator setpoints, 6 process variables, 3 valve position feedbacks, and 6 other control variables collectively capturing the full sensing actuation 
loop of the reactor. The network ($M_3$) modality encodes per-packet 
Modbus/TCP semantics with function code, register address, and raw 
payload alongside flow-identity tuples and epoch timestamps, enabling both command-level and timing-based analysis. In addition, The raw network capture modality retains the complete 
\texttt{.pcap} binary record of all traffic traversing the OT network 
interface, preserving full packet payloads, link-layer headers, and 
per-packet timestamps at libpcap resolution. The host and PLC provenance modalities ($M_1$ and $M_2$) represent system activity as 
directed W3C-PROV-compatible~\cite{missier2013w3c} graphs of process and artifact vertices connected by syscall-level causal edges.

Table~\ref{tab:distribution} details the scale of the collection.  The 48-hour benign phase yields over 4 million host provenance nodes, 28 million host edges, and 11 million Modbus packets, providing a substantial baseline for anomaly detectors to learn normal behavior.  The 22-hour attack phase spans four campaigns that exercise 13 distinct ATT\&CK tactics, producing an additional ${\sim}$ 2 million host nodes and ${\sim}$ 15 million host edges.  Although the total collection duration (70 hours) is shorter than some benchmarks (e.g., HAI~\cite{shin2020hai} at 30 days), the per-hour data density is substantially higher due to the four-modality design. A single hour of our collection produces host provenance graphs, PLC provenance graphs, decoded Modbus logs, raw packets, and physical state rows simultaneously which no prior dataset achieves, making each hour of data far more informative for multi-modal and cross-layer detection research.

Table~\ref{tab:dataset_comparison} compares our dataset with existing relevant datasets. Existing CPS datasets typically capture one or two observation planes.  SWaT~\cite{mathur2016swat}, WADI~\cite{ahmed2017wadi}, and HAI~\cite{shin2020hai} record physical sensor/actuator state but omit host-level and network telemetry entirely, limiting detectors to the physical process view.  ICSSIM~\cite{dehlaghi2023icssim} adds raw PCAP and decoded ICS logs but provides no host provenance.  CICAPT-IIoT~\cite{ghiasvand2024cicapt} contributes provenance graphs yet lacks file and network data-flow edges (i.e., \texttt{read}/\texttt{write}/\texttt{sendto}/\texttt{recvfrom} operations are essentially absent) and multi-host capture.  DARPA TC~\cite{griffith2020scalable} offers the richest provenance among prior works but targets an enterprise-only IT environment.

Our dataset contains four time-synchronized modalities, $M_1$, $M_2$, $M_3$, and $M_4$. The combination of these properties positions ProvICS to support detection approaches that are infeasible with existing datasets. Multi-host provenance ($M_1$, $M_2$) captures cross-boundary anomalies as attackers pivot between the supervisory host and PLC. Synchronized protocol-semantic network ($M_3$) and physical-process data ($M_4$) support multimodal fusion by linking Modbus behavior, process-state deviations, and provenance context. ATT\&CK ICS labels further enable tactic-level evaluation across kill-chain stages rather than only aggregate scoring.

\section{Baseline Intrusion Detection Evaluation}

\begin{figure}[t]
\centering
\includegraphics[width=\linewidth]{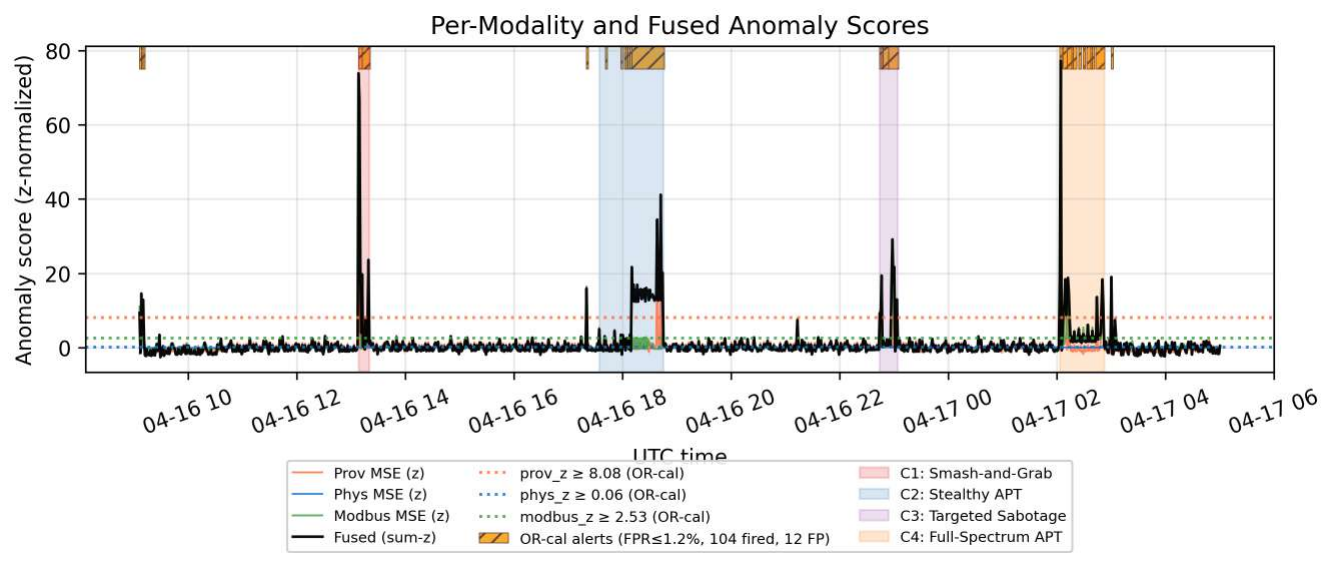}
\caption{Per-modality and fused anomaly-score timeline across the four attack campaigns.}
\label{fig:baseline_scores}
\vspace{-1.6em}
\end{figure}

We validate the ProvICS dataset's detection tractability using three benign trained autoencoders: a GraphSAGE autoencoder~\cite{hamilton2017inductive} over the host and PLC provenance modalities ($M_1$ and $M_2$), a three-layer MLP autoencoder~\cite{feng2018autoencoder} over 94-dimensional windowed physical process features ($M_4$), and another GraphSAGE~\cite{hamilton2017inductive} graph autoencoder over per-window Modbus semantic graphs ($M_3$). Each modality produces a per-window reconstruction error, which is $z$ normalized against the benign distribution and evaluated using late fusion. We use event-level evaluation: each labeled attack phase is treated as one event and is counted as detected if at least one anomalous 60\,s window overlaps its time interval; alerts outside labeled attack phases are counted as false positives and remain penalized. As shown in Table~\ref{tab:baseline_event_detection}, no single modality detects all 32 attack phases: provenance detects 24, physical process detects 17, and Modbus detects 22. In contrast, three-modality sum-$z$ fusion detects all 32 phases, achieving 100\% event-level recall and 0.9133 event-level F1 at a benign window false-positive rate (FPR) of 1.40\%. We also evaluate an FPR constrained OR fusion rule, where each modality has an independently calibrated anomaly threshold and an alert is raised if any modality exceeds its threshold. This stricter OR calibrated fusion setting reduces the FPR to 1.31\% while still detecting 29 of 32 phases. Figure~\ref{fig:baseline_scores} shows the temporal complementarity of the modalities, where different attack phases activate different provenance, physical process, and Modbus anomaly signals.

\begin{table}[t]
\centering
\caption{Baseline event-level detection results for 32 labeled attack phases across four campaigns.}
\label{tab:baseline_event_detection}
\scriptsize
\setlength{\tabcolsep}{5pt}
\renewcommand{\arraystretch}{0.95}
\begin{tabular}{lrrrrrr}
\toprule
\textbf{Detector} & \textbf{Events} & \textbf{TP} & \textbf{FN} & \textbf{Recall} & \textbf{F1} & \textbf{FPR} \\
\midrule
Provenance only         & 32 & 24 & 8  & 0.7500 & 0.6818 & 0.0140 \\
Physical only           & 32 & 17 & 15 & 0.5312 & 0.6358 & 0.0140 \\
Modbus only             & 32 & 22 & 10 & 0.6875 & 0.7556 & 0.0140 \\
Max-$z$ fusion          & 32 & 31 & 1  & 0.9688 & 0.8915 & 0.0140 \\
\textbf{Sum-$z$ fusion} & \textbf{32} & \textbf{32} & \textbf{0} & \textbf{1.0000} & \textbf{0.9133} & \textbf{0.0140} \\
Max3-$z$ fusion         & 32 & 32 & 0  & 1.0000 & 0.9153 & 0.0140 \\
OR-calibrated fusion    & 32 & 29 & 3  & 0.9062 & 0.8892 & 0.0131 \\
\bottomrule
\end{tabular}
\vspace{-2.4em}
\end{table}

\begin{table}[t]
\centering
\caption{Comparison with Existing ICS/CPS Intrusion Detection Datasets}
\label{tab:dataset_comparison}
\scriptsize
\setlength{\tabcolsep}{2pt}
\renewcommand{\arraystretch}{0.95}
\resizebox{\columnwidth}{!}{%
\begin{tabular}{@{}lccccccc@{}}
\toprule
\textbf{Capability} &
\textbf{DARPA~\cite{griffith2020scalable}} &
\textbf{CICAPT~\cite{ghiasvand2024cicapt}} &
\textbf{SWaT~\cite{mathur2016swat}} &
\textbf{WADI\cite{ahmed2017wadi}} &
\textbf{HAI~\cite{shin2020hai}} &
\textbf{ICSSIM~\cite{dehlaghi2023icssim}} &
\textbf{ProvICS} \\
\midrule
\multicolumn{8}{l}{\textit{Host Provenance}} \\
Process lifecycle          & \checkmark & \checkmark & $\times$ & $\times$ & $\times$ & $\times$ & \checkmark \\
File data flow             & \checkmark & $\times$ & $\times$ & $\times$ & $\times$ & $\times$ & \checkmark \\
Network data flow          & \checkmark & $\times$ & $\times$ & $\times$ & $\times$ & $\times$ & \checkmark \\
Event-loop coverage        & Partial    & $\times$ & $\times$ & $\times$ & $\times$ & $\times$ & \checkmark \\
Multi-host provenance      & \checkmark & $\times$ & $\times$ & $\times$ & $\times$ & $\times$ & \checkmark \\
\midrule
\multicolumn{8}{l}{\textit{Network}} \\
Raw PCAP                   & Partial    & \checkmark$^\dagger$ & $\times$ & $\times$ & $\times$ & \checkmark & \checkmark \\
Decoded ICS logs           & $\times$ & $\times$ & \checkmark & $\times$ & $\times$ & \checkmark & \checkmark \\
\midrule
\multicolumn{8}{l}{\textit{Physical Process}} \\
Sensor/actuator state      & $\times$ & $\times$ & \checkmark & \checkmark & \checkmark & \checkmark & \checkmark \\
Real PLC hardware          & $\times$ & Partial    & \checkmark & \checkmark & \checkmark & $\times$ & \checkmark \\
\midrule
\multicolumn{8}{l}{\textit{Ground Truth \& Labeling}} \\
Attack labels              & \checkmark & \checkmark & \checkmark & \checkmark & \checkmark & \checkmark & \checkmark \\
ATT\&CK ICS mapping        & $\times$ & $\times$ & $\times$ & $\times$ & $\times$ & $\times$ & \checkmark \\
Cross-modal bridges        & $\times$ & $\times$ & $\times$ & $\times$ & $\times$ & $\times$ & \checkmark \\
\midrule
Physical Process           & Enterprise & IIoT sim. & Water & Water & HIL & Generic & CSTR \\
Duration                   & 8-14 d & 168 h & 11 d & 16 d & 30 d & Var. &70 h \\
\bottomrule
\end{tabular}%
}
\vspace{0.2em}
\parbox{\columnwidth}{\scriptsize
$^\dagger$Simulated via NS-3. \\
\checkmark~=~present; $\times$~=~absent or not applicable. Dataset duration: 48-hour benign phase + 22-hour attack phase across four campaigns.}
\vspace{-3.0em}
\end{table}

\begin{table}[t]
\centering
\caption{Data distribution across phases and attack tactics}
\label{tab:distribution}
\scriptsize
\setlength{\tabcolsep}{2pt}
\renewcommand{\arraystretch}{0.95}
\resizebox{\columnwidth}{!}{%
\begin{tabular}{@{} l l r r r r r r @{}}
\toprule
\textbf{Phase} & \textbf{Type} & \textbf{H-N} & \textbf{H-E} & \textbf{P-N} & \textbf{P-E} & \textbf{MB} & \textbf{Phys} \\
\midrule
Phase 1 & Benign & 4,113,492 & 16,006,282 & 7,174 & 12,952,594 & 11,798,001 & 125,375 \\
\midrule
Phase 2 & Benign & 1,448,932 & 6,337,905 & 27,130 & 9,113,900 & 4,806,622 & 44,823 \\
 & Discovery & 6,643 & 26,269 & 1,688 & 20,804 & 18,448 & 171 \\
 & Lateral Move. & 596 & 1,405 & 344 & 1,721 & 682 & 6 \\
 & Impair Proc. Ctrl. & 44,388 & 189,108 & 456 & 149,145 & 142,403 & 1,333 \\
 & Inhibit Resp. Func. & 8,191 & 36,798 & 456 & 33,384 & 28,239 & 244 \\
 & Restore & 1,822 & 8,489 & 12 & 6,828 & 6,736 & 65 \\
 & Collection & 4,386 & 18,069 & 58 & 7,244 & 14,043 & 136 \\
 & Impact & 21,508 & 92,558 & 74 & 47,870 & 69,499 & 595 \\
 & Def. Evasion & 14,039 & 62,353 & 145 & 16,829 & 47,075 & 393 \\
 & Init. Access & 2,027 & 7,246 & 33 & 5,502 & 5,328 & 49 \\
 & Persistence & 858 & 3,454 & 34 & 2,285 & 2,372 & 20 \\
 & Execution & 404 & 1,273 & 72 & 1,043 & 838 & 5 \\
 & Cred. Access & 318 & 674 & 272 & 986 & 749 & 6 \\
 & Withdraw & 31 & 105 & 5 & 3 & 51 & 1 \\
 & Dwell & 41,614 & 175,939 & 488 & 5,711 & 133,056 & 1,232 \\
\bottomrule
\end{tabular}%
}
\vspace{0.2em}
\parbox{\columnwidth}{\scriptsize H-N: Host nodes; H-E: Host edges; P-N: PLC host nodes; P-E: PLC host edges; MB: Modbus packets; Phys: Physical-state.}
\vspace{-1.5em}
\end{table}

\section{Conclusion and Future Directions}
This paper presented a multimodal, provenance-aware CPS intrusion detection dataset (ProvICS) collected from a hardware-in-the-loop ICS testbed following the Purdue architecture. The dataset includes host provenance, PLC-edge provenance, decoded Modbus records with raw PCAP, and physical-process telemetry, all aligned on a common UTC timeline to support cross-modal causal analysis. ProvICS contains a 48-hour benign phase and a 22-hour attack phase spanning four heterogeneous adversarial campaigns, with 32 attack events covering 20 unique ICS ATT\&CK techniques across 37 labeled technique-campaign pairs. Baseline evaluation with benign-trained autoencoders confirms the dataset's effectiveness. No single modality detected all 32 labeled attack events, while three-modality sum-$z$ fusion detected all phases with 100\% event-level recall and 0.9133 F1 at a benign-window false-positive rate of 1.40\%. This demonstrates that the dataset contains complementary, temporally aligned signals for evaluating multimodal OT/PIDS methods. To the best of our knowledge, ProvICS is among the first CPS intrusion detection datasets to jointly provide multi-host kernel-level provenance ($M_1$ and $M_2$), protocol semantic capture ($M_3$), physical process state telemetry ($M_4$), ATT\&CK ICS-mapped ground truth with cross-modal bridges, and real PLC hardware in the loop. The dataset is open source and publicly released on Hugging Face.

The current single-PLC, digital-twin testbed is representative rather than large-scale. Future work will focus on real-time PIDS for CPS, multi-PLC and multi-host scaling, real physical plant integration beyond the Node-RED simulator, and support for wireless ICS protocols such as WirelessHART~\cite{song2008wirelesshart} and encrypted industrial traffic.

\bibliographystyle{IEEEtran}
\bibliography{cas-refs}
\end{document}